\documentclass[a4paper,10pt,aps,preprintnumbers,twocolumn,scriptaddress
]{revtex4-1}

\usepackage{graphicx}
\usepackage{subfig}
\usepackage{amsmath}
\usepackage{amsfonts}
\usepackage{bm}
\usepackage[colorlinks,
]{hyperref}
\hypersetup
{
    colorlinks,%
    citecolor=red,%
    linkcolor=blue,%
	pdfauthor={Tong ZHANG},
	pdftitle={Generating polarization controllable FELs at Dalian coherent light source}
}
\usepackage{caption}
\begin{document}

\preprint{IPAC 2013,WEODB102}

\title{Generating polarization controllable FELs at Dalian coherent light source}

\author{T. Zhang}
\author{H.X. Deng}\email{denghaixiao@sinap.ac.cn}%
\author{D. Wang}
\author{Z.T. Zhao}
\affiliation{Shanghai Institute of Applied Physics, Chinese Academy of Sciences, Shanghai 201800, China}

\author{W.Q. Zhang}
\author{G.R. Wu}
\author{D.X. Dai}
\author{X.M. Yang}
\affiliation{Dalian Institute of Chemical Physics, Chinese Academy of Sciences, Dalian 116023, China}

\date{\today}

\begin{abstract}
  The property of the FEL polarization is of great importance to the user community. FEL pulses with ultra-high intensity and flexible polarization control ability will absolutely open up new scientific realms. In this paper, several polarization control approaches are presented to investigate the great potential on Dalian coherent light source, which is a government-approved novel FEL user facility with the capability of wavelength continuously tunable in the EUV regime of $50-150$ nm. The numerical simulations show that both circularly polarized FELs with highly modulating frequency and $100$ $\mathrm{\mu}$J level pulse energy could be generated at Dalian coherent light source.
\end{abstract}

\maketitle

\section{Introduction}

The coherent x-ray light sources with unprecedent brilliance have been already been achieved by the human beings now, i.e. the hard x-ray free-electron laser (FEL) facilities operating on self-amplified spontaneous emission (SASE) mode, LCLS~\cite{Emma_2010_LCLS_NP} and SACLA~\cite{Ishikawa_2012_SACLA_NP}. To date, the pursuit of much more compact design~\cite{Deng_2012_HarmonicXFELO_PRL,Huang_2012_TGU_PRL}, fully coherency~\cite{Zhao_2012_SDUVEEHG_NP,Amann_2012_LCLSself-seeding_NP} and flexible polarization control~\cite{Kim_2000_crossed_NIMA,Geng_2010_crossed_NIMA} are becoming more and more demanding in the scientific community. Taking the great advantages of FELs, the polarization controllable powerful radiation pulses could be used as efficient probes to investigate the chirality compounds, from infrared to hard x-ray regimes.

Recently, a novel fully coherent EUV light sources have been funded and put into construction at Dalian Institute of Chemical Physics, i.e. Dalian coherent light source (DCLS)~\cite{Zhang_2012_DCLS_IPAC2012,Deng_DCLS_arxiv}. With the capability of generating $100$ $\mathrm{\mu}$J FEL pulses at arbitrary frequency between $50-150$ nm, DCLS also has the great potential on lasing the polarization controllable FELs.

In this paper, we will discuss the polarization control approches at DCLS by the newly built one-dimensional time-dependent FEL simulation code ---\textsc{Pelican}~\cite{Zhang_2013_DCLSpolar_CPC}, also accompanies the outline of DCLS and the polarization control experiment by crossed planar undulators at Shanghai deep ultra-violet free-electron laser test facility (SDUV-FEL)~\cite{Zhao_2004_SDUV_NIMA,Zhang_2012_SDUVpolar_NIMA}.

\section{Dalian coherent light source}

DCLS is designed to be working on the HGHG-FEL principle at $2-5$ harmonics of the optical parametric amplification (OPA) seed laser, accordingly gap-tunable undulators are adjusted to lase at EUV wavelength regime of $50$ to $150$ nm, with the FEL pulse energy surpassing $100$ $\mathrm{\mu}$J. The overview parameters of DCLS can be found from Table~\ref{tab:DCLSparam}. Up to now, DCLS has been approved to step into the construction phase, and the first lasing is expected at the end of 2015.
\begin{table}[h]
    \setlength\tabcolsep{4pt}
    \centering
	\caption{Parameters overview of DCLS}
    \label{tab:DCLSparam}
	\begin{tabular}{llll}
		\hline\hline
		Parameter                 & Symbol                    & Value         & Unit              \\
		\hline
		Electron Beam Energy      & $E_b$                     & $300$         & $\mathrm{MeV}$    \\
		Relative Energy Spread    & $\eta$                    & $1\times10^{-4}$ &                \\
		Norm. Emittance           & $\epsilon_n$              & $1-2$         & $\mathrm{{\mu}m}$ \\
		Peak Current              & $I_{pk}$                  & $300$         & $\mathrm{A}$      \\
		Seedlaser Wavelength      & $\lambda_\mathrm{seed}$   & $240-360$     & $\mathrm{nm}$     \\
		Seedlaser Width (FWHM)    & $\tau_\mathrm{seed}$      & $1.0$         & $\mathrm{ps}$     \\
		Radiator Period length    & $\lambda_r$               & $30$          & $\mathrm{mm}$     \\
		Radiator Parameter        & $a_r$                     & $0.3-1.6$     &                   \\
		FEL Wavelength            & $\lambda_\mathrm{FEL}$    & $50-150$      & $\mathrm{nm}$     \\
		FEL Pulse Energy          & $W_\mathrm{FEL}$          & $\geq 100$    & $\mathrm{{\mu}J}$ \\
		\hline\hline
	\end{tabular}
\end{table}

\section{FEL Polarization Control}

Conventionally, there are two approaches to directly control the polarization status of FEL pulses, i.e. by the elliptical permanent undulator (EPU) and the crossed planar undulator (CPU). CPU approach was originally proposed by Kim in high-gain FEL~\cite{Kim_2000_crossed_NIMA} and further extended~\cite{Geng_2010_crossed_NIMA,Ding_2008_crossed_PRSTAB,Li_2010_crossed_NIMA}, from which two identical planar undulators with orthogonal magnetic orientations are responsible for the two linear polarized components of the final received FEL pulses, the electro-magnetic phase shifter between the two undulators is used to tuning the phase differences so as to achieving the modulated polarization.

While the EPU approach is much more straightforward, the two crossed fields are generated simultaneously but the polarization cannot be tuned fast due to the slow mechanical movement. However, FELs from EPU could be almost perfect circularly polarized, which have already been demonstrated at the Elettra storage ring~\cite{Spezzani_2011_ElettraEPU_PRL} and FERMI FEL~\cite{Allaria_2012_FermiFEL_NP}.

In order to take the imperfection effects of EPU, we develop \textsc{Pelican} to study the FEL polarization control at DCLS both with CPU and EPU approaches.

\subsection{Crossed Planar Undulator}

\begin{figure}[h]
  \centering
  \includegraphics[width=1\columnwidth]{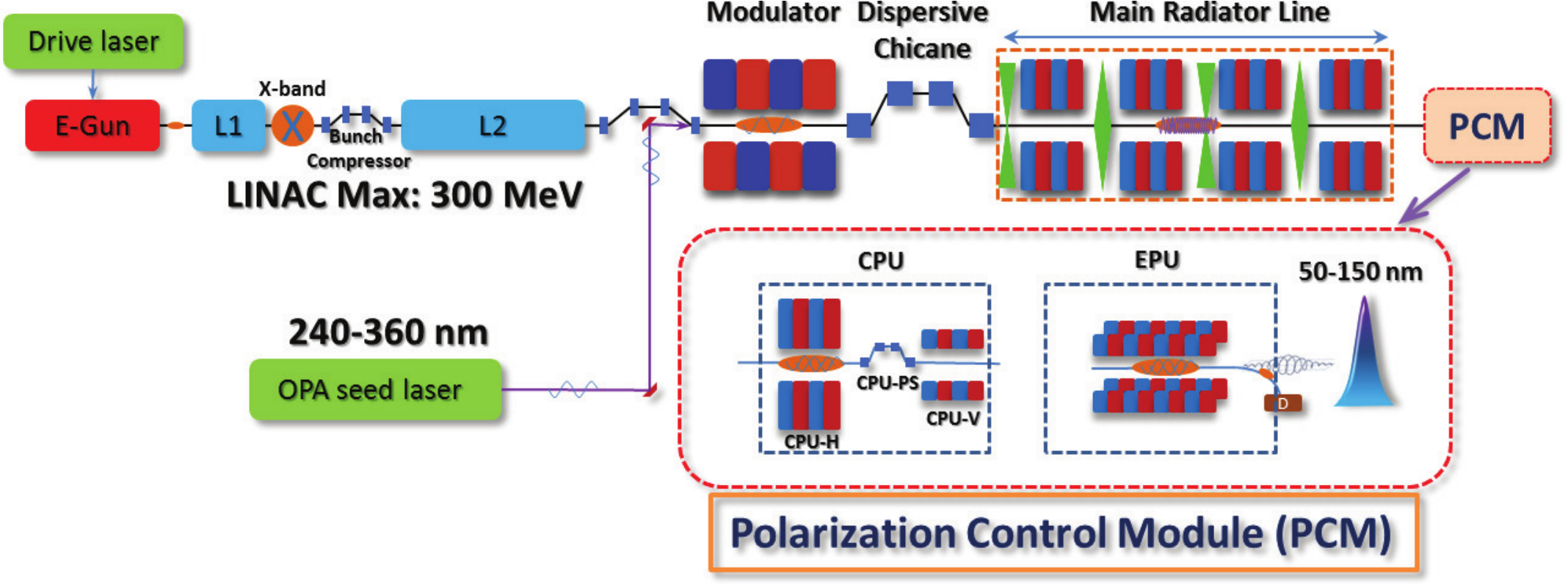}\\
  \caption{Schematic layout of DCLS with polarization control module.}
  \label{fig:layout}
\end{figure}

The schematic layout of DCLS is shown in Fig.~\ref{fig:layout}, from which one can also see the polarization control module (PCM). To control the FEL polarization, PCM could be used as CPU or EPU, respectively. The CPU approaches could be used as when the gaps of the main radiator line are opened up, i.e. the micro-bunched electron beam drifts through to the end where a small section of CPU is placed. The CPU is grouped by two planar undulator with the same total length of 1.5 m and period of 30 mm.

\begin{figure}[h]
  \centering
  \subfloat[PolarAll][Polarizations]{
    \includegraphics[width=0.475\columnwidth]{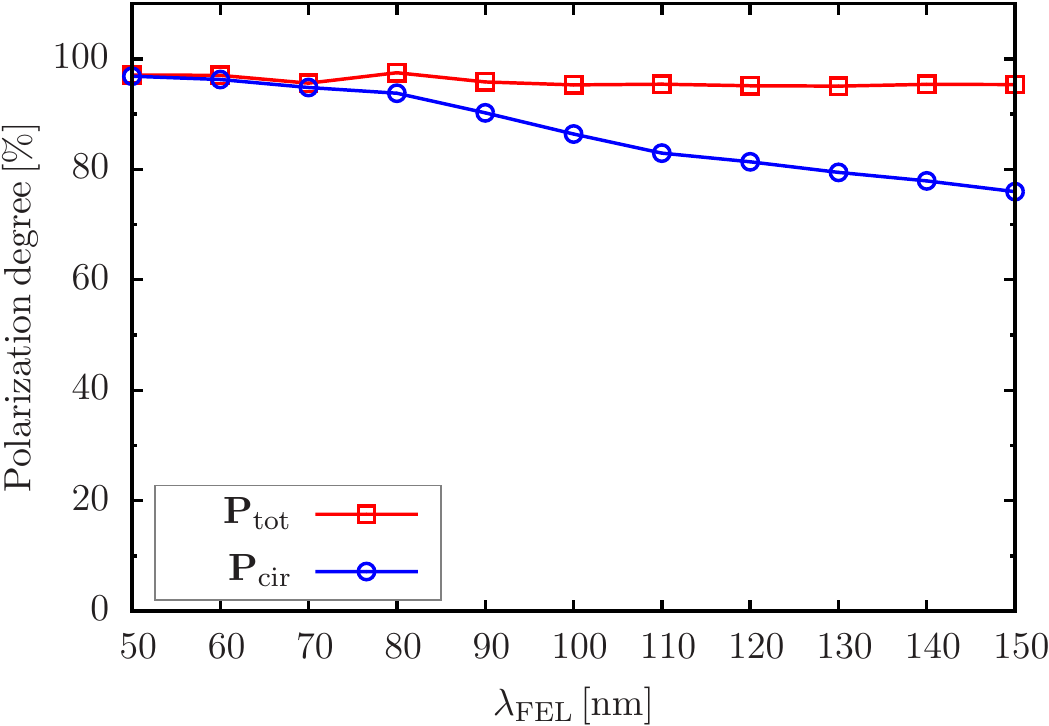}
    \label{subfig:polarAllnoX}
  }
  \subfloat[Polar50ps][Polatization switch]{
    \includegraphics[width=0.475\columnwidth]{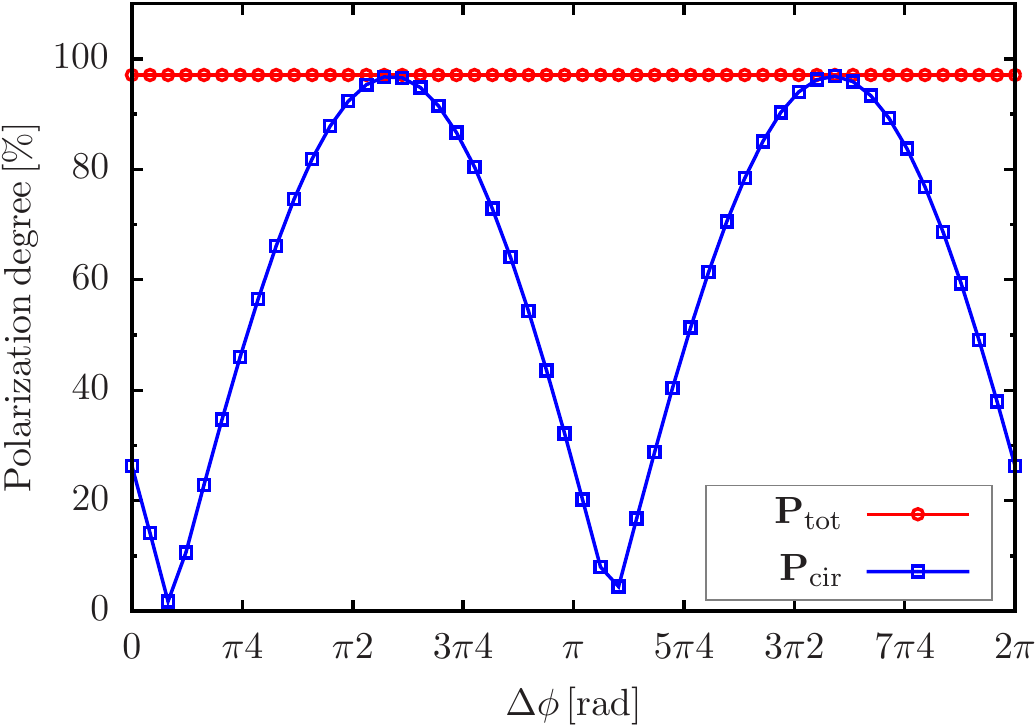}
    \label{subfig:polar50ps}
  }
  \caption{S2E simulations, \emph{left}: polarization degrees of the FEL pulses with CPU (X-band off),
                            \emph{right}: polarization modulating by phase-shifter.}
  \label{fig:polarnoX}
\end{figure}

The start-to-end (s2e) simulation is performed with \textsc{Astra}, \textsc{Elegant}, \textsc{Genesis} and \textsc{Pelican}. Consider cases of electron beam with X-band structure or not, the simulated results indicate that the slippage effect together with the nonlinear residual energy chirp contribute less influences to the final polarized FEL pulses with X-band. The typical polarization degrees over the full operational wavelength regime could be found Fig.~\ref{subfig:polarAllnoX}, and Fig.~\ref{subfig:polar50ps} shows the highly circularly polarized FELs ($\lambda_s=50$ nm) could be obtained with the handedness switching between left and right.

CPU also can be used another manner, i.e. keep the main radiator line still, let the radiated electron beam pass through CPU to achieve much high FEL power, but it is hard to control for achieving FEL pulses with much higher circular polarization, due to the rather different power gain obtained in the two undulators.

\subsection{Elliptical Permanent Undulator}

\begin{figure}[h]
  \centering
  \includegraphics[width=0.7\columnwidth]{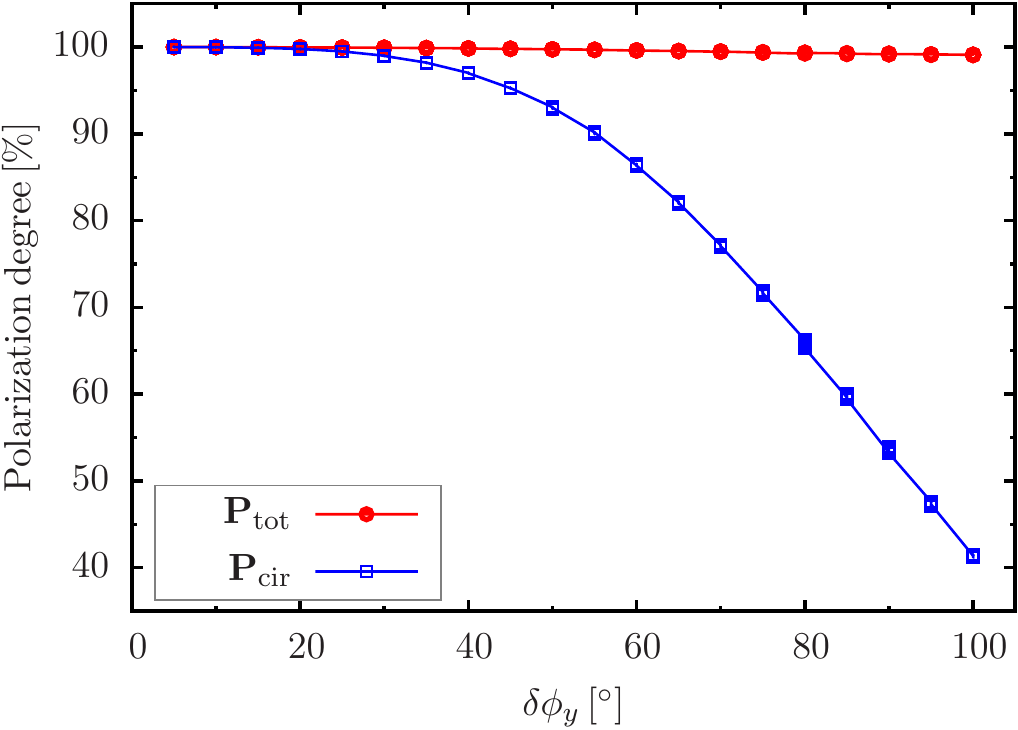}
  \caption{Phase error effect to the FEL polarization in EPU}
  \label{fig:EPU50err}
\end{figure}

Since the FEL fields, $E_x$ and $E_y$ are always simultaneous in the EPU, thus the polarization control approaches mentioned in CPU are also suitable. Here we use \textsc{Pelican} to investigate the undulator phase error effect to the polarization property from EPU. The EPU of PCM has the total length of 3 m and the same period length as CPU's. Firstly, the electron beam traverse the opening main radiator line to radiate. The phase error is taken into account only at one direction, e.g. $\delta\phi_x=0$, $\delta\phi_y$ varies from $0-100^{\circ}$, the simulated polarization status could from Fig.~\ref{fig:EPU50err}, so it is optimistic that the phase error within $20^{\circ}$ could make almost perfect circularly polarized FEL pulses from EPU.

On the other hand, the EPU approah could also be used to generate high circularly polarized FELs with powerful pulse energy output. After undergoing proper exponential gain in the main radiator line, the generation of powerful nearly perfect circularly polarized FELs relies on the fact that the FEL power increases several times in the last placed EPU section.

\begin{figure}[h]
  \centering
  \subfloat[AEPU power][Power]{
    \includegraphics[width=0.48\columnwidth]{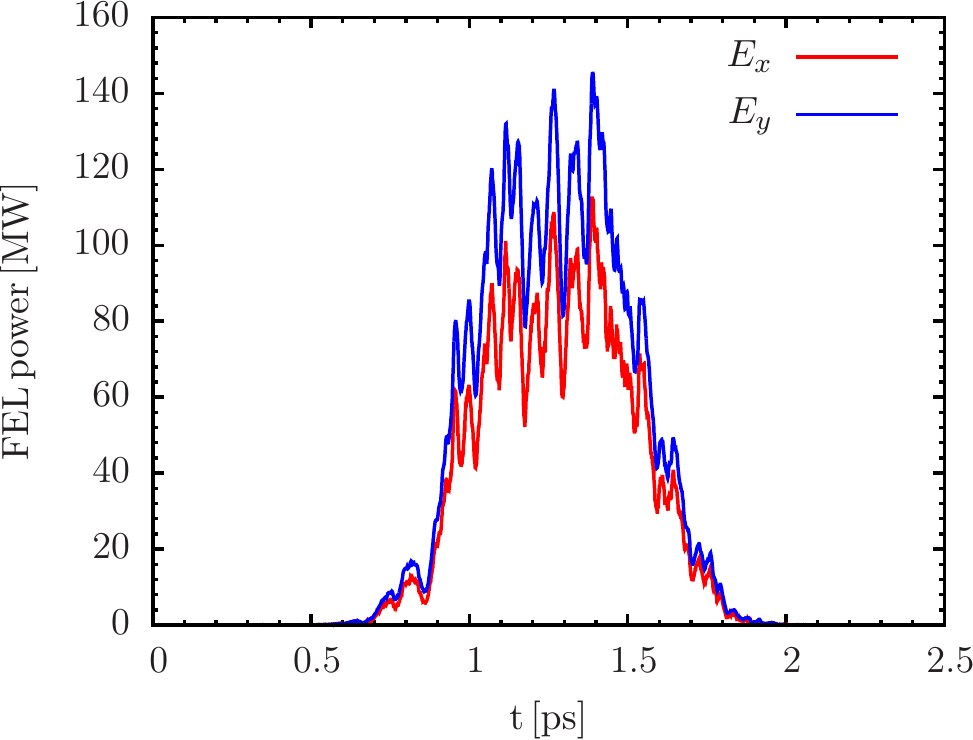}
    \label{subfig:AEPU50power}
  }
  \subfloat[AEPU phase][Phase difference]{
    \includegraphics[width=0.47\columnwidth]{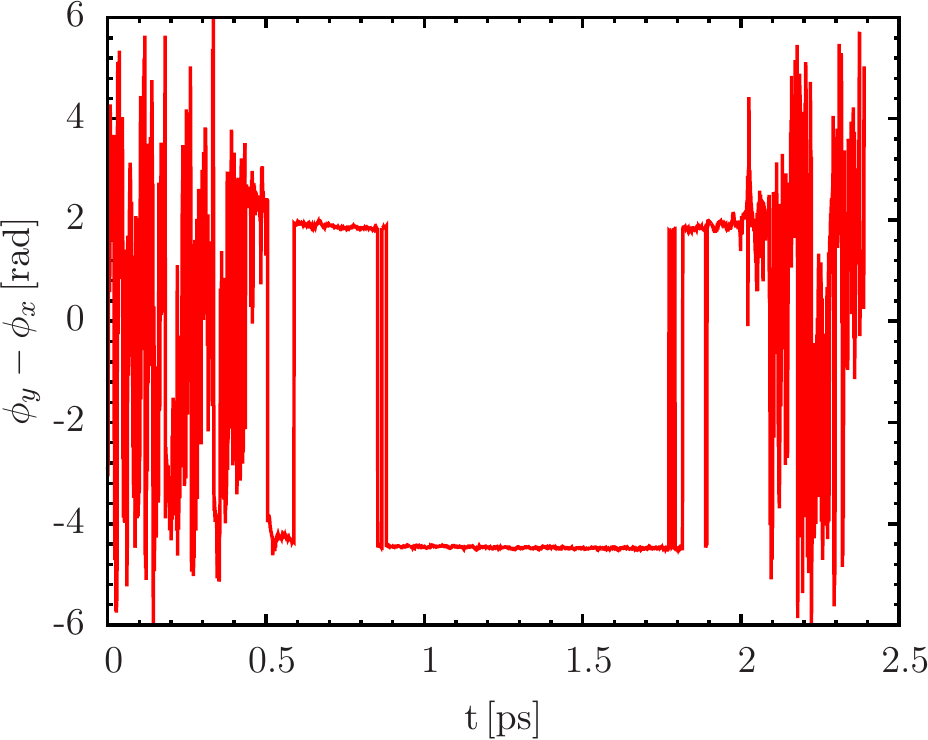}
    \label{subfig:AEPU50phase}
  }
  \caption{S2E simulations for circular polarization FEL with high power output,
                \emph{left}: FEL powers in $E_x$ and $E_y$ direction,
                \emph{right}: The phase difference of $E_x$ and $E_y$.}
  \label{fig:AEPU50pulse}
\end{figure}

\begin{figure}[h]
  \centering
  \includegraphics[width=0.7\columnwidth]{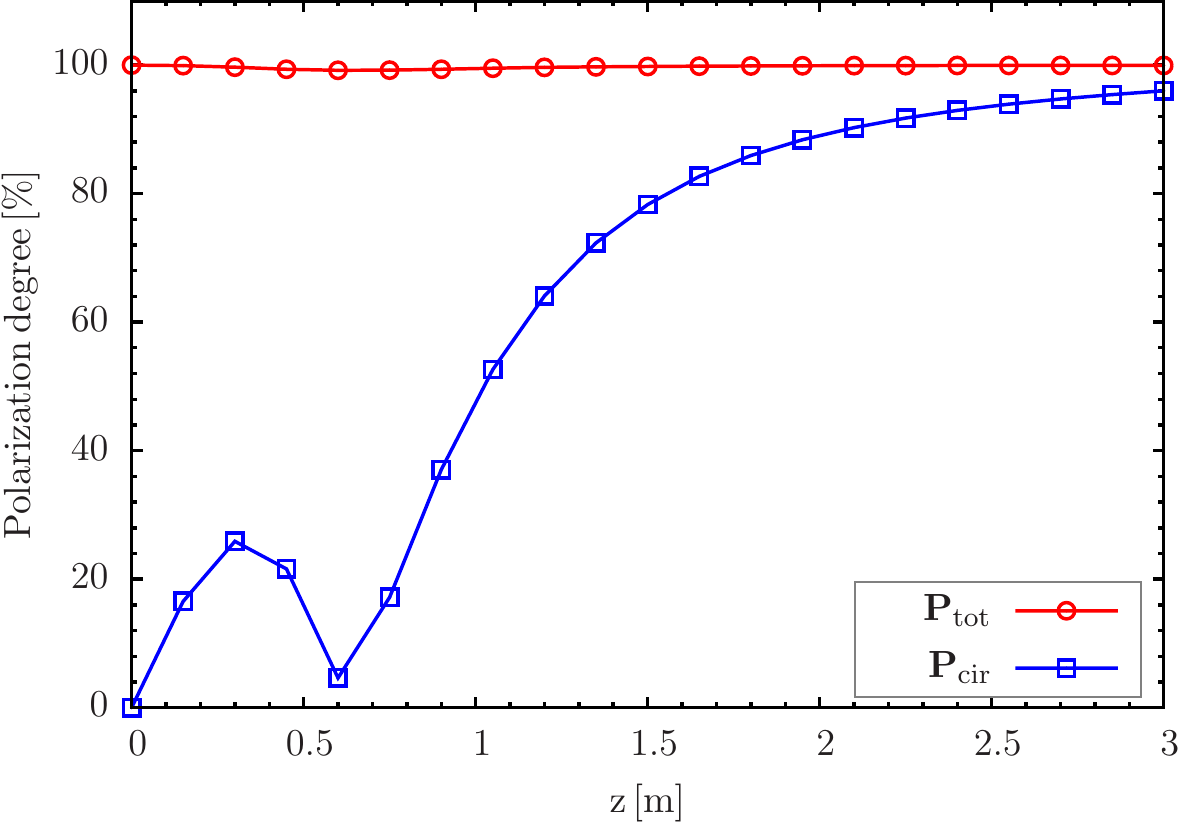}
  \caption{The polarization status evolution along the EPU section.}
  \label{fig:AEPU50polar}
\end{figure}

Here we simulate the case at the wavelength of $50$ nm. The radiated bunched electron bunch contributed linear polarized FEL pulse with the peak power of $\sim$ $30$ MW is then sent into the last EPU to stimulate FEL gain at the crossed orientation to produce $E_y$ and simultaneously amplify $E_x$. Due to the densely micro-bunched electron beam, $E_x$ and $E_y$ grows significantly within about $1-2$ power gain length, i.e. the nearly same level field amplitudes and smooth phase difference could be formed. Fig.~\ref{fig:AEPU50pulse} shows the FEL pulses at the end of EPU module, the polarization evolution along the EPU section could be found from Fig.~\ref{fig:AEPU50polar}. One can see that about $95\%$ circularly polarized FEL pulse is obtained; and the pulse energy is about $130$ $\mathrm{\mu}$J while the output energy is only $2$ $\mathrm{\mu}$J in the case with only one EPU section is used.

\section{CPU experiment at SDUV-FEL}

The FEL polarization control experiment by CPU approach at SDUV-FEL is on-going now~\cite{Deng_2012_SDUVpolar_FEL2012}. Since this is a proof-of-principle experiment for the future Shanghai soft x-ray FEL facility, the learnt experiences could also give some clues for the polarization control at DCLS~\cite{Zhang_2012_SDUVpolar_NIMA}.

\begin{figure}[h]
  \centering
  \includegraphics[width=1\columnwidth]{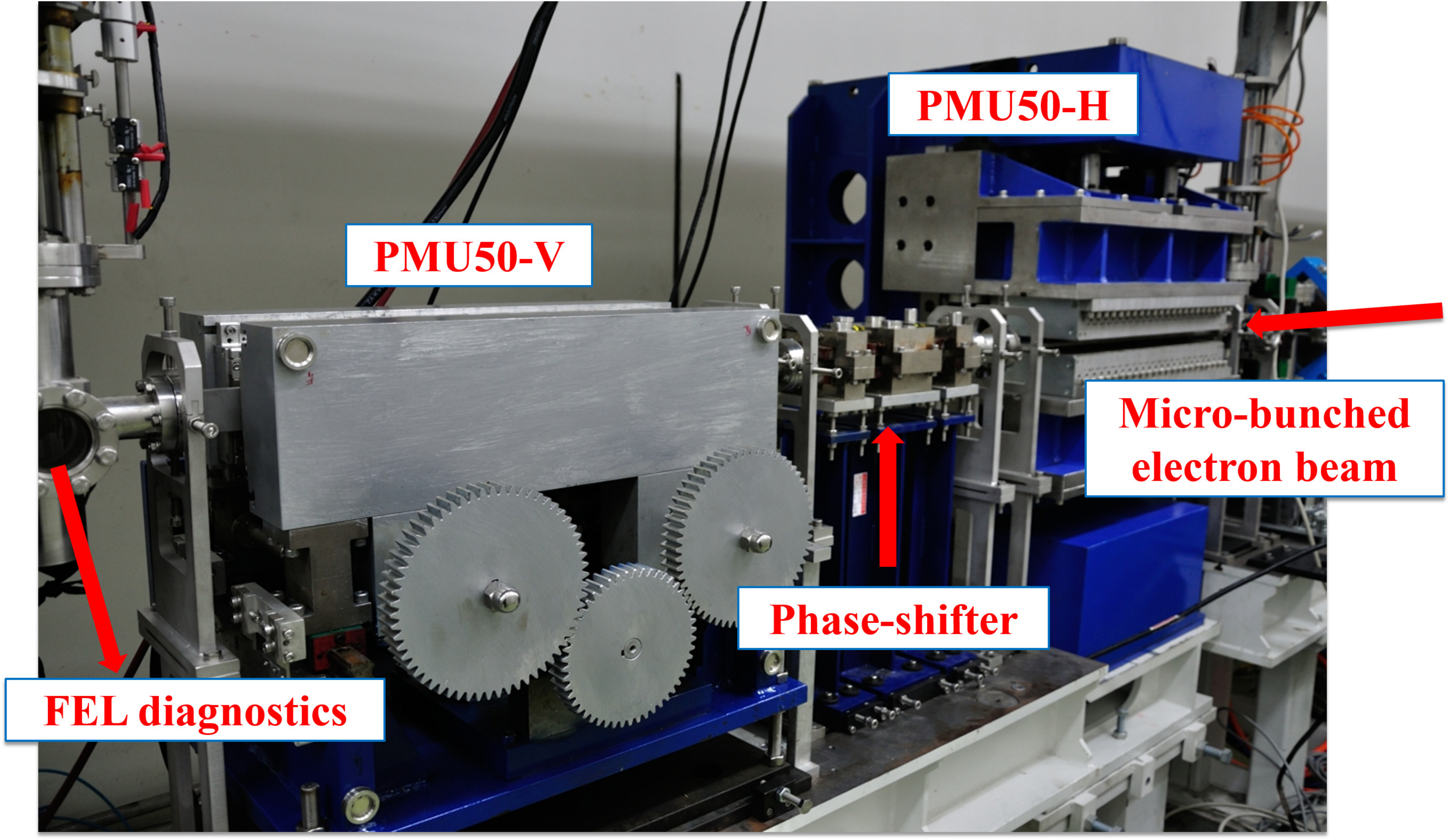}
  \caption{The installed CPU with phase-shifter at SDUV-FEL.}
  \label{fig:SDUV_CPU}
\end{figure}

\begin{figure}[h]
  \centering
  \subfloat[PMU50-V][Trajectory]{
    \includegraphics[width=0.5\columnwidth]{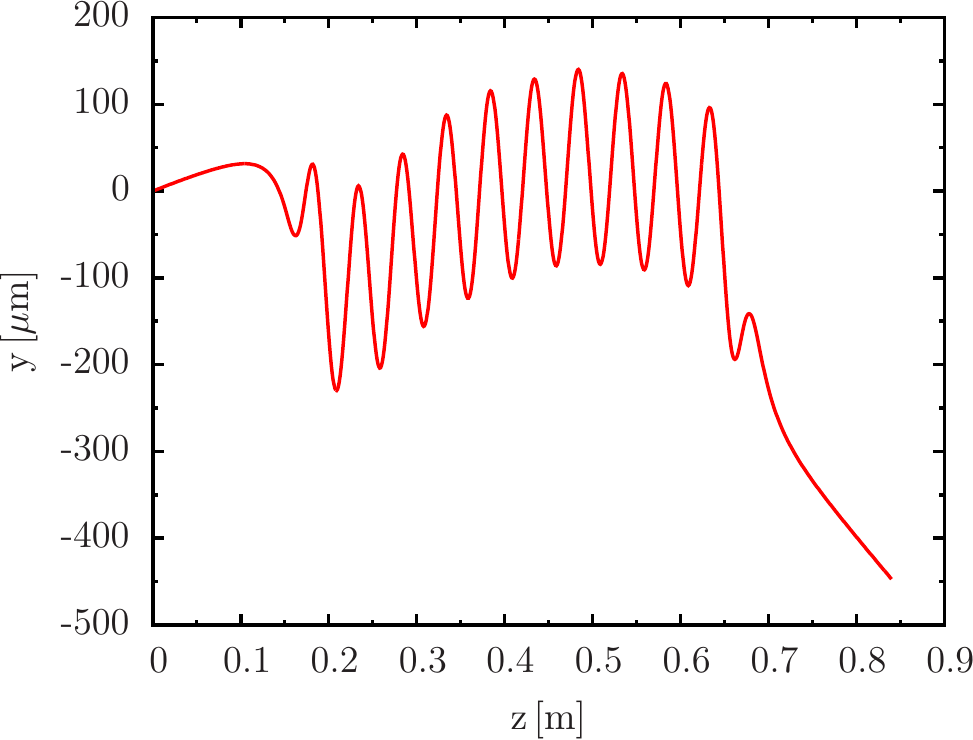}
    \label{subfig:trajPMU50V}
  }
  \subfloat[CPU ps][Phase shifter]{
    \includegraphics[width=0.5\columnwidth]{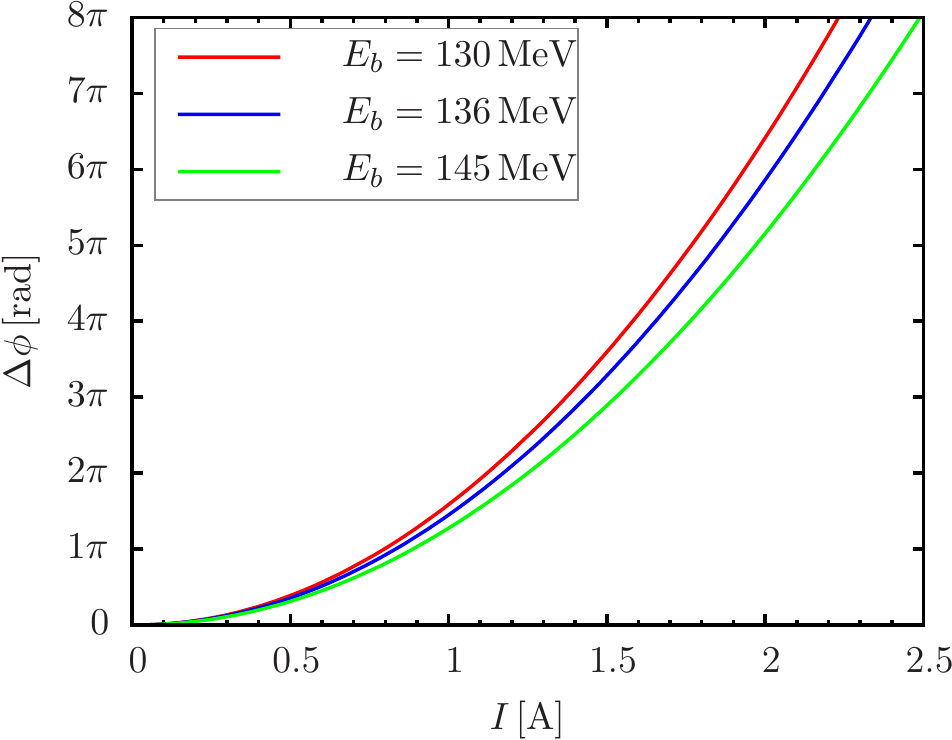}
    \label{subfig:shiftedphase}
  }
  \caption{Trajectory and shifted phase from magnetic measured result of PMU50-V and phase-shifter,
            \emph{left}: calculated electron trajectory in PMU50-V at the gap of 14 mm,
            \emph{right}: the shifted phases at different stimulated currents.}
  \label{fig:trajnphase}
\end{figure}

Now the CPU with electro-magnetic phase-shifter section are installed on the SDUV-FEL now, Fig.~\ref{fig:SDUV_CPU} shows the photograph of such configuration. The magnetic measurements of PMU50-V (which is a newly built undulator to produce the vertically linear polarized FELs) and the small phase-shifter indicate the hardware fulfill the requirement of the physical designs. The electron trajectory and shifted phases could be found from Fig.~\ref{subfig:trajPMU50V} and Fig.~\ref{subfig:shiftedphase}, respectively. The off-line optical diagnostic station already has been setup and calibrated, the next plan is to install them into the final optical measurement hutch and configure the remote control system~\cite{SDUV_CPU}. Experimental results are expected to be got later this summer if all goes well.

\section{Conclusions}

The polarization control of FEL pulses is becoming more and more demanding in the user community. In this paper, we have proposed several FEL polarization control approaches for Dalian coherent light source, the s2e simulations with newly developed FEL code indicate that FEL pulses with flexible polarization characteristics could be achieved. By the crossed planar undulator with phase shifter, DCLS could provide FEL pulses with fast switching polarization, but with small amount pulse energy, while by proper configuring the elliptical permanent undulator, DCLS could serve the user with nearly circularly polarized FEL pulses with the pulse energy surpass $100$ $\mathrm{\mu}$J. Also the polarization control plans at DCLS will benefit a lot from the on-going CPU polarization experiment on high-gain seeded FEL at SDUV-FEL. It could envision that the polarization control with CPU and EPU or even a pair of crossed EPU will make the light sources from free-electron laser facilities much more useful and versatile.

\begin{acknowledgments}
    This work is supported by Major State Basic Research Development Program of China under grant No. 2011CB808300, and Natural Science Foundation of China under grant No. 11175240 \& 11205234.
\end{acknowledgments}

\end{document}